# Predicting Student's Performance Through Data Mining

Aaditya Bhusal
University of Northampton
Northampton, Northamptonshire, UK

**Abstract**: Predicting the performance of students early and as accurately as possible is one of the biggest challenges of educational institutions. Analyzing the performance of students early can help in finding the strengths and weakness of students and help the perform better in examinations. Using machine learning the student's performance can be predicted with the help of students' data collected from Learning Management Systems (LMS). The data collected from LMSs can provide insights about student's behavior that will result in good or bad performance in examinations which then can be studied and used in helping students performing poorly in examinations to perform better.

1. Introduction

Predicting students' performance is one of the biggest challenges faced not only by most of the schools, colleges, and universities but also by the academic sector in general. Educational institutes want to analyze the performance of their students, identify their strengths and weaknesses, and help them perform better in their examinations. The usage of Learning Management System (LMS) is growing among institutions to manage the learning content provided by them along with managing the activities and observing the behaviors of their students while using such systems. Analyzing the behavior of students in LMSs can be beneficial not only in getting the feedback about the quality of the learning content provided to the students but it can also benefit institutions in predicting the performance of students based on their past grades and activities performed by them in the system.

This paper discusses the use of data mining and machine learning methods to help educational institutes to analyze their students' data and predict the performance of their students so that they could provide any required help to the students that are at the risk of scoring low marks at their examinations. (Aljarah, 2016) The project created for this paper uses the data of 480 students of a university collected from a Learning Management System (LMS) called Kalboard 360. Kalboard 360 is a multi-agent LMS designed to facilitate learning through the use of leading-edge technology. The dataset contains a total of 16 features of the students such as their gender, nationality, raised hands, topic chosen etc. The students are classified into 3 classes based on the grades scored by them. The machine learning model trained on this dataset predicts the class of the students bases on the other features present in the dataset.

The report consists of the background and problem space that contains the background information of the research and the problem space the paper is working on. The related work section contains the description of the research done in the similar problem space of student performance prediction. The prototype section contains a detailed description of how the machine learning model was trained. It includes a detailed description of the dataset used for training the machine learning model along with the visualization of some important features related to the students in the dataset. It will also include the description of the data encoding and preprocessing part of the program. Finally, the decisions taken for training the model will be discussed along with the accuracy of the model on both training and testing sets.

## 2. Literature Review

### 2.1. Background and problem space

The competition between the educational institutes is increasing with every year. Every school, college or university wants their students to give their best performance in every examination. The results achieved by the students are not only valuable and pivotal to the students but also to the institutes they study in. With the rise of e-learning technologies it has become easier for educational institutes to provide valuable learning resources to their students. These LMSs can also provide valuable data and insights regarding the usage of the system and behavior of students. They can provide information related to how active the student is while taking a course with the help of the parameters such as how many times the student has interacted with the content of the course, how many times the students have taken quizzes and tests in the LMS, how active the student is while watching an educational video or a textual content in the LMS. These LMS software can give deep insights about the behavior of students which would not be possible to achieve through regular interaction with the student.

The data collected through LMSs can be used to observe the behavior and performance of the student scoring both high and low marks. The behaviors can be studied using LMS software that corelate with the behaviors of the student who performs better in examination. The learnings taken from the behavior of students who have scored higher grades can be used to help the students scoring lower grades in improving their performance in examinations. This could not only help students understand what they are doing wrong but also provide them tips and techniques that they can use to improve their performance in examinations. The data can also help the institute to incorporate intervention strategies into the learning techniques of the students with lower grades at an early stage of their curriculum so that the students could improve their performance by the end of the curriculum. The intervention strategies that are implemented right in the early stage can also boost the confidence of the student not only in their examination but also in their overall academic journey.

### 2.2. Related Work

**Predicting Student Performance by Using Data Mining Methods for Classification [1]**
This paper presents the results from a data mining research project implemented at a Bulgarian university. The project's main goal was to find out the potential of data mining applications for university management and efficient enrollment of the desirable students. The research was conducted in four phases which were Business Understanding Phase, Data Understanding Phase, Data Preprocessing Phase and Modeling Phase. The classifies used for the testing process were Decision tree classifier, Bayesian classifiers and the k-Nearest Neighbour Classifier.

The results achieved by the study were not remarkable. The prediction rates varied from 52-67%. During the prediction, the classifiers perform differently for the five classes mentioned in the research paper. However, from the conclusions made from the study the author considered the results from the performed study as the initial steps and will be used to achieve more accurate results and define further progress in this field.

**Student Performance Prediction using Machine Learning [2]**
In this paper, the authors proposed a model in order to predict the performance of students in an academic organization. The study uses Neural Networks to determine which features correlate with students' performance. In the study, the importance of several different attributes, was considered, in order to determine which of the attribute correlated with student performance.

The paper discusses about the algorithms used in the study such as Bayesian classification algorithm and the features of the students in the dataset. The study confirms that student's past performances have significant influence over their current and future performance and also confirms that neural networks' performance increases with increase in the size of the dataset.

**Towards the integration of multiple classifier pertaining to the Student's performance prediction [3]**
This paper proposes the notion of the integrated multiple classifiers for predicting the students' academic performance. The integrated multiple classifiers consist of algorithms such as Decision Tree, K-Nearest Neighbor, and Aggregating One-Dependence Estimators. The classifier provides a generalized solution for student performance prediction by employing a product of probability combining rule on three student performance datasets.

The study presents a heterogeneous multiple classifier-based framework that integrates multiple classifiers and proposes a single composite model that was tested on three different student datasets. Though the model predicts the performance of engineering students, it can be applicable for other domains.

**The potential for student performance prediction in small cohorts with minimal available attributes [4]**

This paper presents an experiment conducted on a small student cohort of a final-year university module. The data of individual students consisted of attendance, virtual learning environment accesses and intermediate assessments.

The authors found that the predictions about the marks of small student cohorts with very limited attributes would support module leaders to identify the students at risk of failing their assessments. The study suggests a variety of possible interventions that could be done by module leaders at different assessments to improve students' performance.

**Centralized student performance prediction in large courses based on low-cost variables in an institutional context [5]**

This paper presents a prediction model based on low-cost variables and a sophisticated algorithm for predicting early the risk of failing a course by students who are attending large classes. It aims at enabling instructors to carry out early interventions to prevent course failure by using Learning Management System (LMS) data.

This study proposes a low-cost model for predicting students' performance which does not require active effort for data collection. It also presented an efficient Random Forest algorithm which was further improved by LMS data.

**Using learning analytics to identify successful learners in a blended learning course [6]**

This paper presents a three-year case study involving 337 students who attended an academic course and used Moodle for learning. The study shows that students' grades correlated with their attitude and perceptions towards using Moodle.

The study used four techniques i.e., data visualization, decision trees, class association rules, and clustering. The study was performed on a specific set of students, lacks variety and does not account for student's important non-cognitive aspects.

**Study on student performance estimation, student progress analysis, and student potential prediction based on data mining [7]**

This paper formulates a student model with performance and non-performance related attributes. The authors provided multiple analysis tools to analyze student performance, progress and potentials in different ways. The study collected data from 60 high school students.

The tools provided in the study help in predicting various attributes such as students' learning abilities, scores, performance, progress and development. The conducted experiment results show that the estimated results related to various student attributes are accurate and can help in understanding an individual student's learning abilities better.

**Prediction of Student Dropout in E-Learning Program Through the Use of Machine Learning Method [8]**
This paper proposes prediction models to predict the dropout potential of students using their personal characteristics and academic performance as input attributions. The models were developed using Artificial Neural Network (ANN), Decision Tree (DT) and Bayesian Networks (BNs). The model was trained and tested on a large dataset of 62, 375 students. Each model's results were presented in a confusion matrix and analyzed by calculating the rates of accuracy, precision, recall etc.

All of the three presented models suggested relatively effective results in predicting the dropout students with the DT model showing better performance. The methods proposed were used in dropout prediction of students in a university in China. The study aims at improving the precision of predictions made by the model.

**Dropout prediction in e-learning courses through the combination of machine learning techniques [9]**
This paper proposed a dropout prediction method based on three popular machine learning techniques for e-learning courses. The study used three different machine learning techniques and tested them in terms of accuracy, sensitivity and precision.
The proposed method achieved a high prediction rate indicating that the model was more accurate in avoiding misclassifications and correctly identifying dropouts than other studies. The proposed method is expected to help instructors in identifying at-risk students and focusing on their needs.

**Early dropout prediction using data mining: a case study with high school students [10]**
In this paper, the authors proposed a methodology and a specific classification algorithm to discover prediction models of student dropout at an early stage. The study collected the data from 419 high school students in Mexico. The authors carried out several experiments to predict dropout at different steps of the course and compared the model with some traditional algorithms.

The results show that the algorithm was able to predict student dropout within 4-6 weeks of the start of the course. The proposed ICRM2 algorithms outperformed all the traditional classification algorithms in terms of the accuracy of predicting dropout students.

# Prototype

The prototype system used in this project is taken from a Jupyter notebook from Kaggle that uses a popular dataset related to students' academic performance from Kaggle [11] . The prototype system uses the dataset and encodes all the non-numeric values to numeric values.

The prototype did not have any visualization of the data in the dataset, so the author added several visualizations that presented the information in the dataset visually and gave some more insights about the data in the dataset.

The prototype system predicts the academic performance of students based on a collection of related attributes. The prototype uses a Tensorflow model that uses the student's dataset for training the model and predicting the grades of the students based on their other related attributes which are included in the dataset. The dataset used for training the machine learning model is collected from a multi-agent Learning Management System (LMS) called Kalboard 360.

## Dataset

The dataset was used on a paper which proposed a new student performance model with a new category of features called student's behavioral features [12, 13]. The data was collected using a learner activity tracker tool called experience API (xAPI). The xAPI is a component of the training and learning architecture (TLA) and helps to determine the learner, activity and objects that describe a learning experience.The dataset consists of 480 student records and was collected through two educational semesters. The dataset consists of a list of features related to the students such as their gender, nationality, place of birth, grade level, raised hands, absent days etc. The features in the dataset are classified into three major categories which are demographic features such as the gender and nationality of the students, academic background related features such as educational stage, grade level and section the student belongs to and the behavioral features of students such as number of times their hands were raised in the class, the number of times they opening the resources in the learning management system, answering survey taken by the parents of the students, and the overall school satisfaction

The dataset consists of a total number of 305 males and 175 females. The students in the dataset come from different origins such as Kuwait, Jordan, Palestine Iraq, Lebanon, Tunis, Saudi Arabia, Egypt, Syria, USA, Iran, Libya, Morocco, and Venezuela. The dataset was collected from the LMS through two educational semesters of the course. In the dataset a total of 245 student records were collected during the first semester of the academic course and a total of 235 student records are collected during the second semester of the academic course. The dataset also consists of other related student features such as visited resources, section, semester, topic, parent satisfaction from school etc. Finally, the students in the dataset are classified into three numerical intervals based on their total grade i.e. low-level (0-69), mid-level (70-89) and high-level (90-100).

```
<class 'pandas.core.frame.DataFrame'>
RangeIndex: 480 entries, 0 to 479
Data columns (total 17 columns):
 #   Column                   Non-Null Count  Dtype
---  ------                   --------------  -----
 0   gender                   480 non-null    object
 1   NationalITy              480 non-null    object
 2   PlaceofBirth             480 non-null    object
 3   StageID                  480 non-null    object
 4   GradeID                  480 non-null    object
 5   SectionID                480 non-null    object
 6   Topic                    480 non-null    object
 7   Semester                 480 non-null    object
 8   Relation                 480 non-null    object
 9   raisedhands              480 non-null    int64
 10  VisITedResources         480 non-null    int64
 11  AnnouncementsView        480 non-null    int64
 12  Discussion               480 non-null    int64
 13  ParentAnsweringSurvey    480 non-null    object
 14  ParentschoolSatisfaction 480 non-null    object
 15  StudentAbsenceDays       480 non-null    object
 16  Class                    480 non-null    object
dtypes: int64(4), object(13)
memory usage: 63.9+ KB
```

*Figure 1 - Number of columns in the dataset*

## Visualization

The following are the visualization of some important features of the students in the dataset. The visualizations were not the part of the code used for this project and were created by the author of this report. The visualization shown were of student's gender, nationality, raised hands on classroom, chosen topic and the class based on their grades.

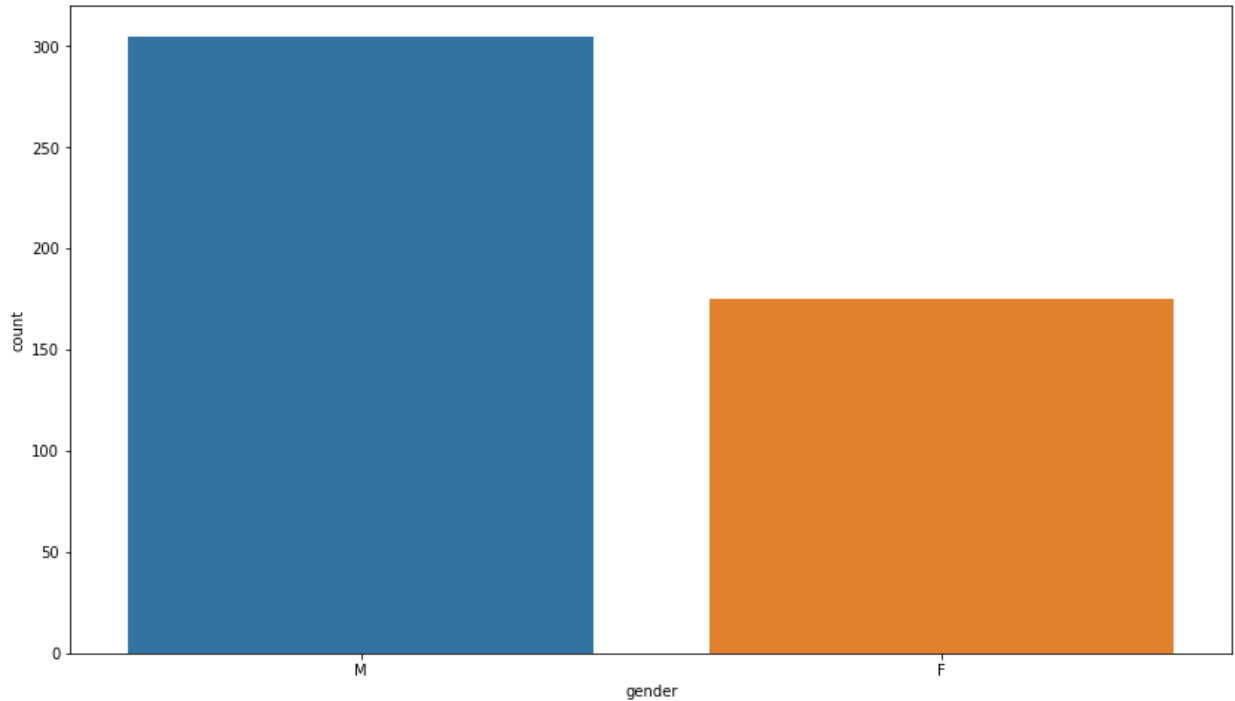

*Figure 2 - Number of males and females in the dataset*

From the visualization we can see that the number for male students represented by blue color in the visualization is about 300 and the number of females represented by blue color in the visualization is about 170 which is almost half the number of total male students.

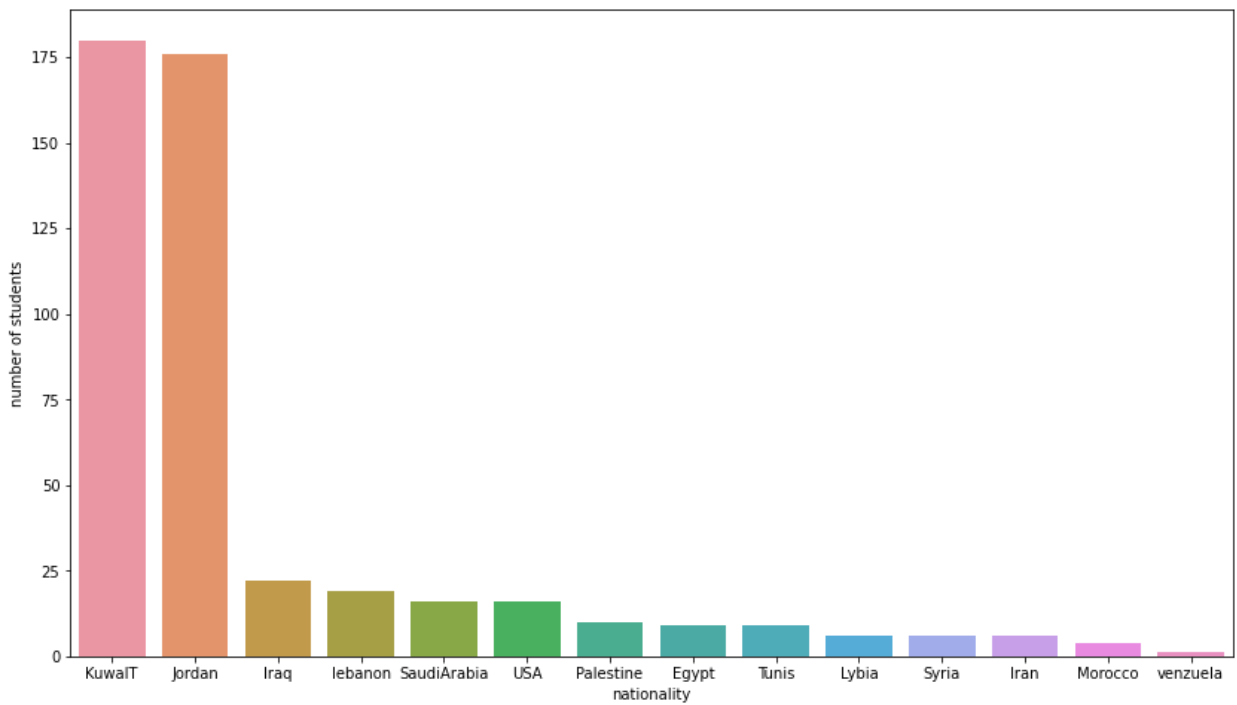

*Figure 3 - Student's nationality in the dataset*

Most of the students in the dataset belong to the nationality of Kuwait and Jordan, both having more than 175 students. The two nationalities are then followed by Iraq, Lebanon, Saudi Arabia, USA, Palestine, Egypt, Tunis, Libya, Syria, Iran, Morocco, and Venezuela.

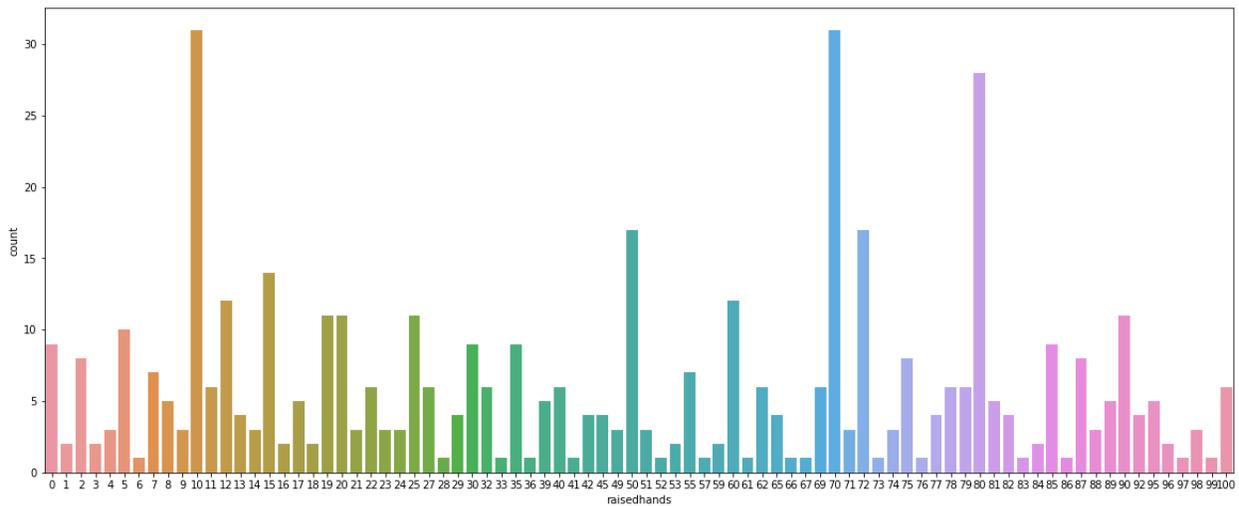

*Figure 4 - Number of hands raised by the students*

The number of times the students raised their hands is distributed roughly throughout the scale of 0 to 100 with the hands being raised mostly 10, 70 and 80 times.

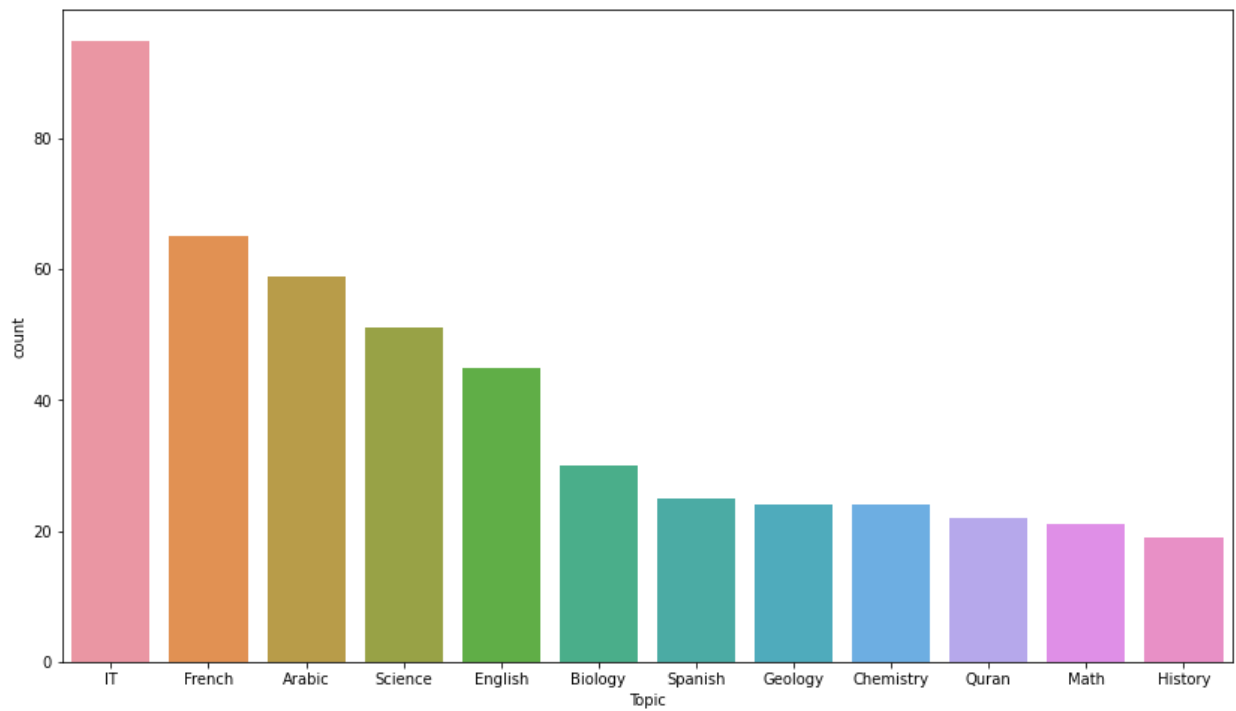

*Figure 5 - Topics chosen by the students*

Most of the students had chosen IT as their topic. Other topics chosen by most students are French, Arabic, Science and English.

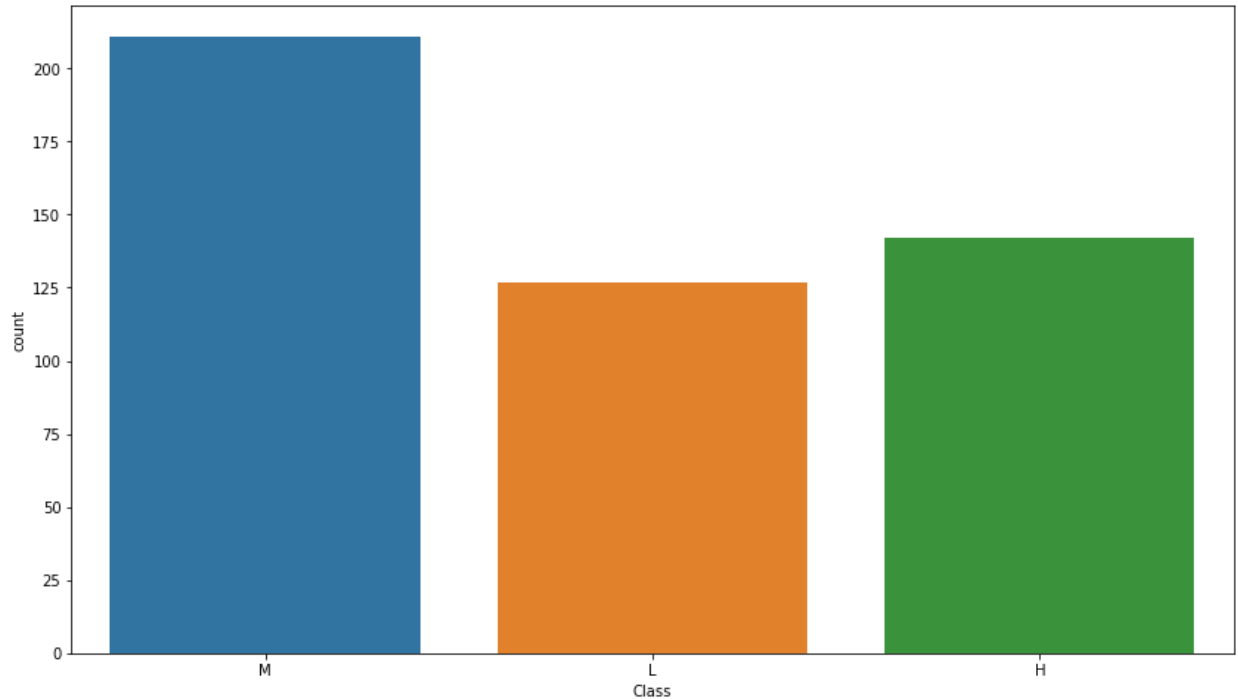

*Figure 6 - Number of students in the Class column*

In the dataset, most of the students have scored 70-89 (middle-level) followed by 90-100 (high level) and 0-69 (low-level)

# Encoding and Preprocessing

The dataset had many columns with non-numerical values which needed to be converted to numerical values. First, the dataset was checked whether it had any null values or not. There were no null values in any of the columns of the dataset. Then, the features of the dataset were divided into three categories i.e., binary features, ordinal features and nominal features. The binary features were the features that only has two values in the entire column of the dataset. The ordinal features were the features that could be placed in a particular order. The nominal features were the features which had multiple values and did not have any form of ordering.

The positive and negative values of all the binary features were differentiated and given a value of 0 and 1. Similarly the ordinal features were given the ordering of their values as their numerical values during encoding. The nominal features were given the prefixes in front of their values in their columns. The values of binary, ordinal and nominal features were converted into their corresponding numerical values using encoding functions. The following images displays that all non-numerical in values in the dataset have been converted into their corresponding numerical values.

|   | gender | StageID | GradeID | Semester | Relation | raisedhands | VisITedResources | AnnouncementsView | Discussion | ParentAnsweringSurvey | ... |
|---|--------|---------|---------|----------|----------|-------------|------------------|-------------------|------------|----------------------|-----|
| 0 | 1 | 0 | 1 | 0 | 1 | 15 | 16 | 2 | 20 | 1 | ... |
| 1 | 1 | 0 | 1 | 0 | 1 | 20 | 20 | 3 | 25 | 1 | ... |
| 2 | 1 | 0 | 1 | 0 | 1 | 10 | 7 | 0 | 30 | 0 | ... |
| 3 | 1 | 0 | 1 | 0 | 1 | 30 | 25 | 5 | 35 | 0 | ... |
| 4 | 1 | 0 | 1 | 0 | 1 | 40 | 50 | 12 | 50 | 0 | ... |
| ... | ... | ... | ... | ... | ... | ... | ... | ... | ... | ... | ... |
| 475 | 0 | 1 | 5 | 1 | 1 | 5 | 4 | 5 | 8 | 0 | ... |
| 476 | 0 | 1 | 5 | 0 | 1 | 50 | 77 | 14 | 28 | 0 | ... |
| 477 | 0 | 1 | 5 | 1 | 1 | 55 | 74 | 25 | 29 | 0 | ... |
| 478 | 0 | 1 | 5 | 0 | 1 | 30 | 17 | 14 | 57 | 0 | ... |
| 479 | 0 | 1 | 5 | 1 | 1 | 35 | 14 | 23 | 62 | 0 | ... |

480 rows × 56 columns

*Figure 7 - Non-numerical values encoded to numerical values*

The dataset was then scaled and split for the machine learning model to train on it. The class column of the dataset which divided the students according to the grades scored by them was removed because it was the data that needed to be predicted based on other features. Then, the StandardScalar function provided by the scikit-learn library was used to scale the data. The data needed to be normalized before training the model so that the values could be brought down to a similar scale. The scaled dataset was then split into training and testing sets. The dataset was then ready for the model to be trained on.

## Training

The training part of the prototype started by the creation of a Tensorflow model [14]. There were a total 55 features in the training set so, the input layer was given an input shape of 55. Then, two hidden layers were added to the model. Both of the hidden layers were a Dense layer with .. and the activation function used was 'ReLU'. Dense layer or fully connected layer connected each input to each output within its layer. (Team, n.d.) ReLU stands for Rectified Linear Unit. It takes the input and returns the maximum among the input and 0 [15]. The output layer is also a Dense layer. It uses softmax activation function because in this model, the data is classified in more than two classes. The following image shows the summary of the model:

```
Model: "model"
_________________________________________________________________
Layer (type)                 Output Shape              Param #
=================================================================
input_1 (InputLayer)         [(None, 55)]              0
_________________________________________________________________
dense (Dense)                (None, 64)                3584
_________________________________________________________________
dense_1 (Dense)              (None, 64)                4160
_________________________________________________________________
dense_2 (Dense)              (None, 3)                 195
=================================================================
Total params: 7,939
Trainable params: 7,939
Non-trainable params: 0
_________________________________________________________________
None
```

*Figure 8 - Summary of the machine learning model*

The model uses 'adam' optimizer which is suitable for large parameters in a neural network. The loss function used is sparse_categorical_crossentropy which calculates the crossentropy loss between the labels and predictions. The training batch size was given 16 and the number of epochs for the training was 30. The model had about 8000 trainable parameters and 4 layers in total.

After the completion of the training part, the training accuracy of the model was 0.94 and the validation accuracy was 0.68. The model performed very well with the training set and fairly well with the validation set. The change in the training accuracy and validation accuracy is shown in the graph below:

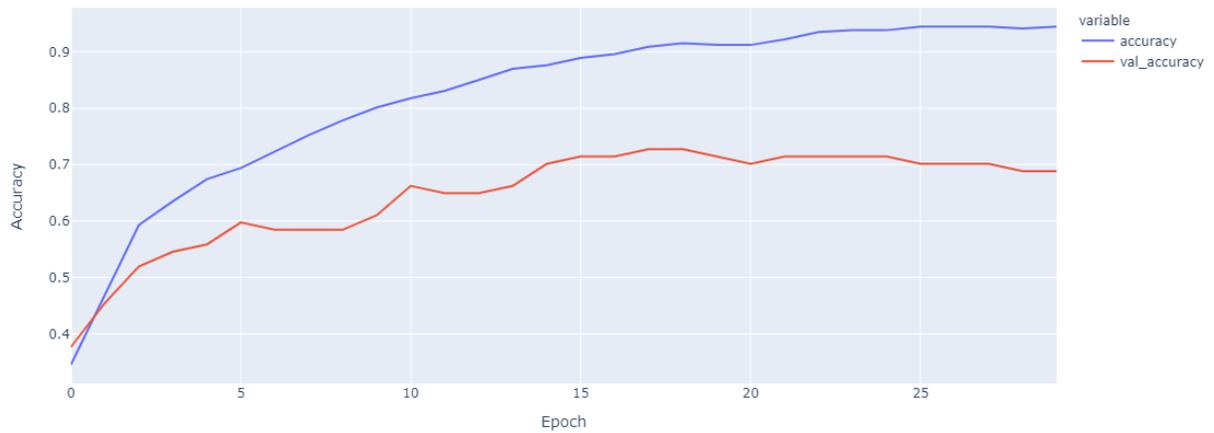

*Figure 9 - Training and Validation Accuracy*

The training loss of the model was 0.22 and the validation loss was 0.75. The model performed better in the training set. The loss in the training set was significantly less than the loss in the validation set. The change in the training loss and validation loss is shown in the graph below:

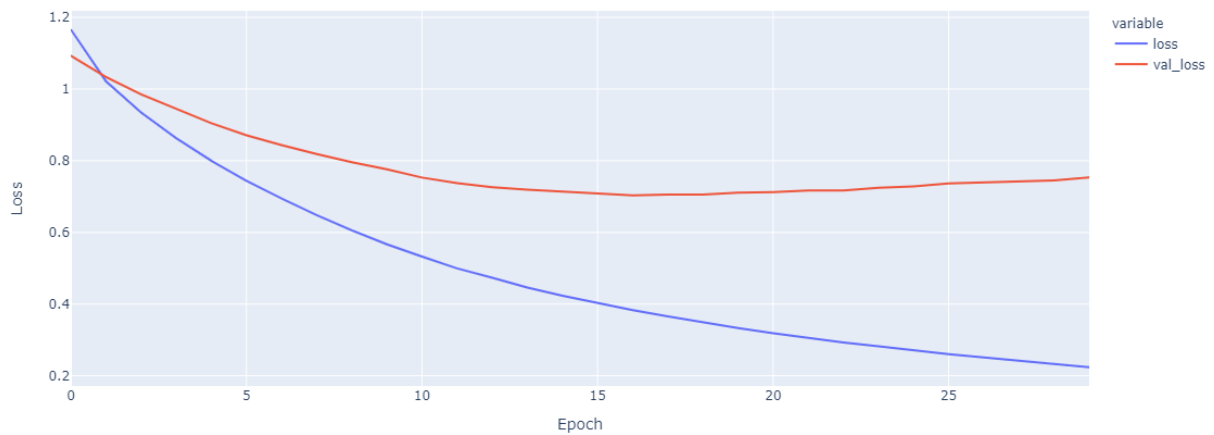

*Figure 10 - Training and Validation loss*

After evaluating the accuracy of the model on the testing data, an accuracy of 0.76 was achieved. This means that the model can perform well in predicting the students' performance based on the other information which were included in the dataset used by the prototype system. The loss and accuracy of the model for the testing dataset is shown in the figure below:

```
3/3 [==============================] - 0s 3ms/step - loss: 0.5611 - accuracy: 0.7604
[0.5611234307289124, 0.76041668865348816]
```

Figure 11 - Model accuracy in testing set

# Conclusions

The author learned many new concepts, techniques, practices and implementations in the field of machine learning throughout the period of doing this project. The field of educational data mining is vast and needs a lot of research and work to bring significant improvement in order to impact the students' performance through the use of various machine learning techniques. While doing the research for this project it was often found that early intervention strategies can cause significant impact in students' performance and also in predicting the dropout rate of the students. It was also found that students' interaction with Learning Management Systems helps the most in predicting the students' final grades in examinations.

The accuracy of the trained model was 0.76 which makes the model good in predicting student's performance. Although the prediction accuracy of the trained model is good, it could have been made better by including the data of more students and also by including more data features of students to train the model. In this report, the author gave an introduction to the project and the field of student performance analysis. The report also contains the background of the project and the problem space the project is working on. The report includes 10 research papers that have conducted studies on similar topics and finally the report contains the prototype system that tries to solve the problem space of the research and a detailed description of how the system was built.